# Moving pattern-based modeling using a new type of interval ARX model

Sun Changping[1, 2] (Email: sunchangping@tjut.edu.cn)

1. Tianjin Key Laboratory for Control Theory & Application in Complicated Systems, Tianjin University of Technology, Tianjin, 300384, China
2. School of Electrical Engineering and Automation, Tianjin University of Technology, Tianjin, 300384, China

**Abstract**: In this paper, firstly, to overcome the shortcoming of traditional ARX model, a new operator between an interval number and a real matrix is defined, and then it is applied to the traditional ARX model to get a new type of structure interval ARX model that can deal with interval data, which is defined as interval ARX model (IARX). Secondly, the IARX model is applied to moving pattern-based modeling. For better characterizing pattern moving space, the "cardinality" of pattern moving space (CPMS) is defined. Finally, to verify the validity of the proposed modeling method, it is applied to a sintering process. The simulation results show the moving pattern-based modeling using the new type of interval ARX model is robust to variation in parameters of the model, and the performance of the modeling using the proposed IARX is superior to that of the previous work.

## 1 Introduction

ARX is a classic model for system modeling, which is widely applied for time series forecasting [1], system modeling [2], fault detection [3], controller design [4] and so on. However, variables in many complex systems in the real world are imprecise, uncertain, and incomplete. For example, when measuring a variable, it is difficult to obtain the exact value of the variable due to the existence of errors. In addition, the price of a stock fluctuates during a given day. From an application point of view, sometimes we don't care what the specific value of a variable is, but if the variable is in some normal or expected range. For example, whether a patient's blood pressure is in the normal range; in a control system, whether a controlled variable is controlled within a certain expected range.

Based on the above considerations, Professor Xu Zhengguang and his research group proposed a modeling method based on moving pattern, and several results have been made. The core idea of moving pattern-based modeling is to forecast the pattern class to which the operating condition pattern belong at the future time according to the pattern classes to which the current and previous operating condition pattern belonged, and then describe the dynamic characteristics of complex production process under the pattern class granularity [5]. According to the different measurement methods of patterns, the current moving pattern-based modeling methods could be divided into three categories. The first category is to use interval numbers or interval vectors to measure the pattern [5-9]. First, the moving pattern space is constructed, and the interval number and interval vector are used to measure the one-dimensional moving pattern and the multidimensional moving pattern, respectively. Then, the corresponding interval dynamic model is constructed, and then the dynamic characteristic is described under the pattern class granularity. An interval ARX model is defined in [7], which predicts interval values of the variable at next moment using central values of several variables at previous moments (whose value are interval numbers), and then a moving pattern-based dynamic model is constructed. In this model structure, the multiplication between interval numbers and real numbers still follows the classical interval arithmetic. The second category is to represent patterns by central values of the classes. For example, the k-means clustering method is used to construct pattern moving space [10-14], and the pattern class variable was measured by the central values of pattern classes in pattern moving space, that is, at some point if the operating condition pattern is classified to some class, then the central value of the class is taken as a measure of the pattern class variable at that time. Maximum posteriori probability is used for measuring the pattern class variable in the third category [15]. Authors in [15] applied maximum posteriori probability to measure the pattern class variable and defined a new type of Petri nets to describe the moving of the pattern.

To better describe the moving of a pattern in the pattern moving space, in this paper, first, a new operator between an interval number and a real matrix is defined. Then, the defined operator is applied to the traditional ARX model. Finally, a new type of interval ARX model capable of processing interval data is obtained and used to construct a moving pattern-based dynamic model.

The structure of the paper is as follows: the second part introduces some basic concepts of moving pattern-based modeling and interval arithmetic used in this paper; the third part introduces a new operator and the new type of interval ARX model proposed in this paper. The fourth part introduces the moving pattern-based modeling using the proposed IARX, and the "cardinality" of pattern moving space (CPMS) is defined. The fifth part is experimental simulation. The sixth part is the conclusion.

## 2 Preliminary knowledge

In this section, some basic concepts of moving pattern-based modeling and interval arithmetic are introduced briefly. For more information, please refer to [5-9][16].

### 2.1 Modeling based on moving pattern [5]

The forecasting model based on moving pattern is as follows:

$$dx(k) = F\big(d\tilde{x}(k)\big)$$

$$= F\left(f\big(dx(k-1), dx(k-2)\cdots dx(k-m)\big)\right) \qquad (1)$$

Where $dx(k)$ is the pattern class variable. $d\tilde{x}(k)$ is the preliminary prediction value of $dx(k)$ predicted by the preliminary prediction model $f(\cdot)$. Here, $F(\cdot)$ denotes classification.

### 2.2 Interval arithmetic [16]

Operations of interval number are as follows. Here, interval numbers $D = [d_l, d_u]$,

$Q = [q_l, q_u]$.

Addition:
$$D + Q = [d_l + q_l, d_u + q_u] \tag{2}$$

Subtraction:
$$D - Q = [d_l - q_u, d_u - q_l] \tag{3}$$

The multiplication between a constant and an interval number:
$$\lambda \cdot D = [\lambda \cdot d_l, \lambda \cdot d_u] \tag{4}$$

Where $\lambda > 0$.
$$\lambda \cdot D = [\lambda \cdot d_u, \lambda \cdot d_l] \tag{5}$$

Where $\lambda < 0$.

If the interval numbers are given in the form of the center and the radius, for example, $D = (d_c, d_r)$, $Q = (q_c, q_r)$, where $d_c = \frac{d_l + d_u}{2}$, $d_r = \frac{d_u - d_l}{2}$, $q_c = \frac{q_l + q_u}{2}$, $q_r = \frac{q_u - q_l}{2}$, then the addition, subtraction, and multiplication between a constant and an interval number are defined as follows:

$$D + Q = (d_c + q_c,\ d_r + q_r) \tag{6}$$
$$D - Q = (d_c - q_c,\ d_r + q_r) \tag{7}$$
$$\lambda \cdot D = (\lambda \cdot d_c,\ |\lambda| \cdot d_r) \tag{8}$$

Where $\lambda$ is a constant.

## 3 A new type of interval ARX model

The traditional ARX model can only be used for system modeling based on crisp data. To overcome the deficiency of traditional ARX model, a new type of interval ARX model is proposed in this paper. Different from the existing interval ARX model based on classical interval arithmetic [17], the proposed new type of interval ARX model is based on the proposed operator between an interval number and a real matrix. The new operator is defined as follows.

### 3.1 Definition of a new operator

The new operator between an interval number and a real matrix is defined as follows:

**Definition 1**: the operator " $\circ$ " between an interval number $(a, c)$ and a $2n \times m$ dimensional real matrix **P** is defined as follows:

$$(a, c) \circ \boldsymbol{P} = (a, c) \circ \begin{bmatrix} p_{11}^1 & p_{12}^1 \cdots p_{1m}^1 \\ p_{11}^2 & p_{12}^2 \cdots p_{1m}^2 \\ p_{21}^1 & p_{22}^1 \cdots p_{2m}^1 \\ p_{21}^2 & p_{22}^2 \cdots p_{2m}^2 \\ \vdots & \vdots & \vdots \\ p_{n1}^1 & p_{n2}^1 \cdots p_{nm}^1 \\ p_{n1}^2 & p_{n2}^2 \cdots p_{nm}^2 \end{bmatrix}$$

$$= \begin{bmatrix} (ap_{11}^1, cp_{11}^2) & (ap_{12}^1, cp_{12}^2) \cdots (ap_{1m}^1, cp_{1m}^2) \\ (ap_{21}^1, cp_{21}^2) & (ap_{22}^1, cp_{22}^2) \cdots (ap_{2m}^1, cp_{2m}^2) \\ \vdots & \vdots & \vdots \\ (ap_{n1}^1, cp_{n1}^2) & (ap_{n2}^1, cp_{n2}^2) \cdots (ap_{nm}^1, cp_{nm}^2) \end{bmatrix} \tag{9}$$

Where, $p_{ij}^2 \geq 0$, $i = 1,2,\cdots,n$, $j = 1,2,\cdots,m$

Particularly, if $m=1$, the formula (9) is simplified as follows:

$$(a, c) \circ \mathbf{P} = (a, c) \circ \begin{bmatrix} p_{11}^1 \\ p_{11}^2 \\ p_{21}^1 \\ p_{21}^2 \\ \vdots \\ p_{n1}^1 \\ p_{n1}^2 \end{bmatrix} = \begin{bmatrix} (ap_{11}^1, cp_{11}^2) \\ (ap_{21}^1, cp_{21}^2) \\ \vdots \\ (ap_{n1}^1, cp_{n1}^2) \end{bmatrix} \quad (10)$$

Where, $p_{i1}^2 \geq 0$, $i = 1,2,\cdots,n$.

### 3.2 The definition of a new type of interval ARX model

Based on the operator defined above, a new type of interval ARX model is proposed.

**Definition 2:** Given a pair of samples $(\tilde{Y}(k), u(k))$, $k = 1,2,\cdots,n$, where $\tilde{Y}(k)$ is an interval output, and $u(k)$ is a crisp input, the new type of interval ARX model is defined as follows:

$$\tilde{Y}(k) = \tilde{p}_0 + \sum_{j=1}^{n} \tilde{Y}(k-j) \circ \mathbf{p}_j + \sum_{l=1}^{m} \tilde{p}_{l+n} u(k-l) \quad (11)$$

Where $\tilde{Y}(k)$ is an interval output, $\tilde{p}_0$ and $\tilde{p}_{l+n}$ are interval parameters, $\mathbf{p}_j$ is a two-dimensional column vector parameter, e.g., $\mathbf{p}_j = \begin{bmatrix} p_j^1 \\ p_j^2 \end{bmatrix}$, $p_j^2$ is a real number, and $p_j^2 \geq 0$, $j = 1,2,\cdots,n$, $l = 1,2,\cdots,m$.

Based on interval arithmetic (see the formula (2)-(8)) and the definition 1, the formula (11) could be further derived as follows:

$$\begin{aligned}
\tilde{Y}(k) &= (p_{0c}, p_{0r}) + \sum_{j=1}^{n} (Y_c(k-j), Y_r(k-j)) \circ \begin{bmatrix} p_j^1 \\ p_j^2 \end{bmatrix} \\
&\quad + \sum_{l=1}^{m} (p_{(l+n)c}, p_{(l+n)r}) \; u(k-l) \\
&= (p_{0c}, p_{0r}) + \sum_{j=1}^{n} (Y_c(k-j) p_j^1, Y_r(k-j) p_j^2) \\
&\quad + \sum_{l=1}^{m} (p_{(l+n)c} u(k-l), p_{(l+n)r} |u(k-l)|)
\end{aligned}$$

$$= \left( p_{0c} + \sum_{j=1}^{n} Y_c(k-j) p_j^1 + \sum_{l=1}^{m} p_{(l+n)c} u(k-l), \quad p_{0r} + \sum_{j=1}^{n} Y_r(k-j) p_j^2 \right.$$

$$\left. + \sum_{l=1}^{m} p_{(l+n)r} |u(k-l)| \right)$$

$$= (\mathbf{A}^T \mathbf{x}(k-1), \mathbf{C}^T |\mathbf{x}(k-1)|) \tag{12}$$

Where

$$\mathbf{A} = [p_{0c}, p_1^1, p_2^1, \cdots p_n^1, p_{(1+n)c}, p_{(2+n)c}, \cdots, p_{(m+n)c}]^T$$

$$\mathbf{C} = [p_{0r}, p_1^2, p_2^2, \cdots, p_n^2, p_{(1+n)r}, p_{(2+n)r}, \cdots, p_{(m+n)r}]^T$$

$$= [c_0, c_1, c_2, \cdots, c_{(m+n)}]^T$$

$$\mathbf{x}(k-1) = [1, Y_c(k-1), Y_c(k-2), \cdots, Y_c(k-n), u(k-1), u(k-2), \cdots, u(k-m)]^T$$

$$|\mathbf{x}(k-1)| = [1, Y_r(k-1), Y_r(k-2), \cdots, Y_r(k-n), |u(k-1)|, |u(k-2)|, \cdots, |u(k-m)|]^T$$

**3.3 Parameter identification in the proposed interval ARX model**

In this paper, the parameters in formula (12) are identified by least square method, i.e.,

$$\min_{\mathbf{A}} \quad J_0 = \sum_{k=1}^{N} \left( Y_c(k) - \mathbf{A}^T \mathbf{x}(k-1) \right)^2 \tag{13}$$

$$\min_{\mathbf{C}} \quad J_1 = \sum_{k=1}^{N} \left( Y_r(k) - \mathbf{C}^T \mathbf{x}(k-1) \right)^2 \tag{14}$$

$$s.t. \ c_j \geq 0, \ j = 0, 1, \ 2, \ \cdots, \ (m+n)$$

Here, $N$ is number of samples. Formula (14) is equivalent to the quadratic programming problem.

**Theorem 1**: The optimal solution of formula (14) is the same as that of formula (15).

$$\min_{\mathbf{C}} \quad J_2 = \mathbf{C}^T \mathbf{H} \mathbf{C} - \mathbf{C}^T \mathbf{B} \tag{15}$$

$$s.t. \ c_j \geq 0, \ j = 0, 1, \ 2, \ \cdots, \ (m+n)$$

Where

$$\mathbf{C} = [p_{0r}, p_1^2, p_2^2, \cdots, p_n^2, p_{(1+n)r}, p_{(2+n)r}, \cdots, p_{(m+n)r}]^T$$

$$= [c_0, c_1, c_2, \cdots, c_{(m+n)}]^T$$

$$\mathbf{H} = \sum_{k=1}^{N} |\mathbf{x}(k-1)| |\mathbf{x}^T(k-1)|, \qquad \mathbf{B} = 2\sum_{k=1}^{N} Y_r(k)|\mathbf{x}(k-1)|$$

$$\mathbf{x}(k-1) = [1, Y_c(k-1), Y_c(k-2), \cdots, Y_c(k-n), u(k-1), u(k-2), \cdots, u(k-m)]^T$$

$$|\mathbf{x}(k-1)| = [1, Y_r(k-1), Y_r(k-2), \cdots, Y_r(k-n), |u(k-1)|, |u(k-2)|, \cdots, |u(k-m)|]^T$$

Proof:

$$J_1 = \sum_{k=1}^{N} (Y_r(k) - \mathbf{C}^T |\mathbf{x}(k-1)|)^2$$

$$= \sum_{k=1}^{N} Y_r^2(k) + \sum_{k=1}^{N} \mathbf{C}^T |\mathbf{x}(k-1)||\mathbf{x}^T(k-1)|\mathbf{C} - 2\sum_{k=1}^{N} Y_r(k)\mathbf{C}^T |\mathbf{x}(k-1)|$$

$$= \sum_{k=1}^{N} Y_r^2(k) + \mathbf{C}^T \left(\sum_{k=1}^{N} |\mathbf{x}(k-1)||\mathbf{x}^T(k-1)|\right)\mathbf{C} - 2\mathbf{C}^T \sum_{k=1}^{N} Y_r(k)|\mathbf{x}(k-1)|$$

$$= \sum_{k=1}^{N} Y_r^2(k) + \mathbf{C}^T \mathbf{H} \mathbf{C} - \mathbf{C}^T \mathbf{B}$$

As $\sum_{k=1}^{N} Y_r^2(k)$ is a constant, so the optimal solution of formula (14) is the same as that of formula (15). □

### 4 Moving pattern-based modeling using the proposed interval ARX

Professor Xu Zhengguang and his research group proposed a modeling method based on moving pattern. The main modeling process refers to [5-9][16].

#### 4.1 Moving pattern-based modeling using the proposed interval ARX model

The proposed interval ARX model is applied to moving pattern-based modeling and the sintering process is used as an example of application. Based on the new type of interval ARX model, the constructed moving pattern-based model is as follows:

$$d\tilde{x}(k) = (p_{0c}, p_{0r}) + \sum_{j=1}^{n} d\,x(k-j) \circ \begin{bmatrix} p_j^1 \\ p_j^2 \end{bmatrix} + \sum_{l=1}^{m} \left(p_{(l+n)c}, p_{(l+n)r}\right) u(k-l) \quad (16)$$

Where $dx(k)$ and $d\tilde{x}(k)$ are the pattern class variable and the preliminary prediction value of the pattern class variable at time k respectively, and u(k) is a crisp input.

Further, $d\tilde{x}(k)$ is written as $\left(d\tilde{x}_c(k),\ d\tilde{x}_r(k)\right)$, and substitute it into the equation (16), formula (17) is derived.
Where

$$d\tilde{x}(k) = \left(d\tilde{x}_c(k), \quad d\tilde{x}_r(r)\right)$$

$$= (p_{0c}, p_{0r}) + \sum_{j=1}^{n} dx(k-j) \circ \begin{bmatrix} p_j^1 \\ p_j^2 \end{bmatrix} + \sum_{l=1}^{m} (p_{(l+n)c}, p_{(l+n)r}) \quad u(k-l)$$

$$= (p_{0c}, p_{0r}) + \sum_{j=1}^{n} (dx_c(k-j), \quad dx_r(k-j)) \circ \begin{bmatrix} p_j^1 \\ p_j^2 \end{bmatrix}$$

$$+ \sum_{l=1}^{m} (p_{(l+n)c}, p_{(l+n)r}) \quad u(k-l)$$

$$= (p_{0c}, p_{0r}) + \sum_{j=1}^{n} (dx_c(k-j) p_j^1, \quad dx_r(k-j) p_j^2)$$

$$+ \sum_{l=1}^{m} (p_{(l+n)c} u(k-l), p_{(l+n)r}|u(k-l)|)$$

$$= \left(p_{0c} + \sum_{j=1}^{n} dx_c(k-j) \; p_j^1 + \sum_{l=1}^{m} p_{(l+n)c} u(k-l), \quad p_{0r} \right.$$

$$\left. + \sum_{j=1}^{n} dx_r(k-j) \; p_j^2 + \sum_{l=1}^{m} p_{(l+n)r}|u(k-l)| \right)$$

$$= (\mathbf{A}^T \mathbf{x}(k-1), \mathbf{C}^T|\mathbf{x}(k-1)|) \qquad (17)$$

$$\mathbf{A} = [p_{0c}, p_1^1, p_2^1, \cdots p_n^1, p_{(1+n)c}, p_{(2+n)c}, \cdots, p_{(m+n)c}]^T$$

$$\mathbf{C} = [p_{0r}, p_1^2, p_2^2, \cdots, p_n^2, p_{(1+n)r}, p_{(2+n)r}, \cdots, p_{(m+n)r}]^T$$

$$= [c_0, c_1, c_2, \cdots, c_{(m+n)}]^T$$

$$\mathbf{x}(k-1) = [1, dx_c(k-1), dx_c(k-2), \cdots, dx_c(k-n), u(k-1), u(k-2), \cdots, u(k-m)]^T$$

$$|\mathbf{x}(k-1)| = [1, dx_r(k-1), dx_r(k-2), \cdots, dx_r(k-n), |u(k-1)|, |u(k-2)|, \cdots, |u(k-m)|]^T$$

### 4.2 Identification of the parameters in the model

Here, the parameters in (17) could be obtained by solving formulas (18) and (19). Here, $N$ is number of samples.

$$\min_{\mathbf{A}} J_3 = \sum_{k=1}^{N} (dx_c(k) - \mathbf{A}^T \mathbf{x}(k-1))^2 \qquad (18)$$

$$\min_{\mathbf{C}} J_4 = \sum_{k=1}^{N} (dx_r(k) - \mathbf{C}^T|\mathbf{x}(k-1)|)^2 \qquad (19)$$

$$s.t. \ c_j \geq 0, j = 0,1,2,\cdots,(m+n)$$

The process of solving equations (18) and (19) is similar to that of equations (13) and (14). The parameters in formula (18) are obtained by least square method. According to theorem 1, the parameters in formula (19) are obtained by solving the formula (15).

**4.3 The final prediction model based on moving pattern**

In the process of modeling based on moving pattern, after the preliminary prediction output $d\tilde{x}(k)$ is obtained, the classification for $d\tilde{x}(k)$ is required, namely $d\tilde{x}(k)$ is classified to a certain class belong to the pattern moving space, and then the interval number representation of this class is used as the measurement value of the final predicted output $d\hat{x}(k)$. For better characterizing pattern moving space, the "cardinality" of pattern moving space is defined as follows.

**Definition 3**: The "cardinality" of pattern moving space is defined as the number of pattern classes constituting the pattern moving space and is denoted as **CPMS**.

In this paper, the Hausdorff distance is used as the measurement of the distance between two interval numbers, and the preliminary predicted output is classified according to the nearest neighbor classification principle.

**4.4 Performance evaluation**

The root mean square errors (RMSE) of the outputs of the preliminary prediction model and the final prediction model are used to quantitatively evaluate the performance of the moving pattern-based modeling using the proposed interval ARX model. The RMSE of the output of the preliminary prediction model is defined as follows:

$$\text{RMSE}_{d\tilde{x}^L} = \sqrt{\frac{1}{N}\sum_{k=1}^{N}(dx^L(k) - d\tilde{x}^L(k))^2} \quad (20)$$

$$\text{RMSE}_{d\tilde{x}^H} = \sqrt{\frac{1}{N}\sum_{k=1}^{N}(dx^H(k) - d\tilde{x}^H(k))^2} \quad (21)$$

Where the upper bound RMSE of the preliminary prediction output are represented by $\text{RMSE}_{d\tilde{x}^H}$. Similarly, $\text{RMSE}_{d\tilde{x}^L}$ denotes the lower bound RMSE. The upper bounds of the pattern class variable $dx(k)$ and the preliminary prediction output $d\tilde{x}(k)$ are represented by $dx^H(k)$ and $d\tilde{x}^H(k)$, respectively. The lower bounds of the pattern class variable $dx(k)$ and the preliminary prediction output $d\tilde{x}(k)$ are represented by $dx^L(k)$ and $d\tilde{x}^L(k)$, respectively.

The RMSE of final prediction output is defined as follows:

$$\text{RMSE}_{d\hat{x}^L} = \sqrt{\frac{1}{N}\sum_{k=1}^{N}(dx^L(k) - d\hat{x}^L(k))^2} \quad (22)$$

$$\text{RMSE}_{d\hat{x}^H} = \sqrt{\frac{1}{N}\sum_{k=1}^{N}(dx^H(k) - d\hat{x}^H(k))^2} \quad (23)$$

Where the upper and lower bound RMSE of the final prediction output are represented by $\text{RMSE}_{d\hat{x}^H}$ and $\text{RMSE}_{d\hat{x}^L}$, respectively. The $d\hat{x}^H(k)$ and $d\hat{x}^L(k)$ represent upper and lower bound of the final prediction output $d\hat{x}(k)$, respectively.

## 5 Simulation and test

For verifying the validity of the proposed moving pattern-based modeling using the new type of interval ARX model, it is applied to the sintering process of a steel plant. The real operating condition data—the exhaust temperature of three bellows—are collected. The sampling period T is 25s, and the number of the samples is 864. After zero-mean normalization and principal component analysis, the dimension reduced operating condition data is obtained, as shown in Fig. 1.

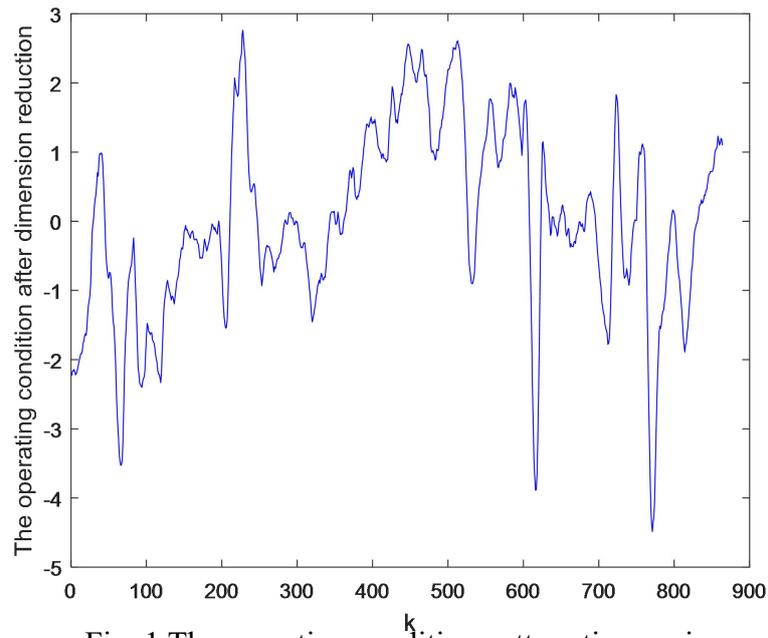

Fig. 1 The operating condition pattern time series

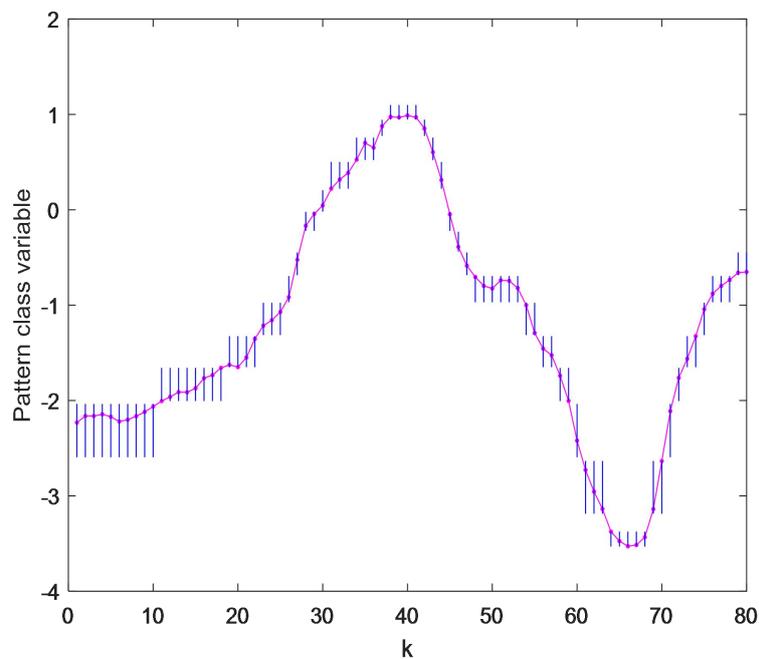

Fig. 2 The pattern class variable time series.

Table I Interval representation of one-dimensional pattern class in pattern moving space

| $P_i$ | $I_i = [I_i^L, I_i^U]$ |
|---|---|
| 1 | [-4.4856, -4.2777] |
| 2 | [-4.1184, -3.8815] |
| 3 | [-3.6970, -3.6746] |
| 4 | [-3.5314, -3.3746] |
| 5 | [-3.1872, -2.6340] |
| 6 | [-2.5955, -2.0353] |
| 7 | [-2.0056, -1.6574] |
| 8 | [-1.6497, -1.3244] |
| 9 | [-1.3137, -0.9751] |
| 10 | [-0.9697, -0.6939] |
| 11 | [-0.6848, -0.4478] |
| 12 | [-0.4387, -0.2304] |
| 13 | [-0.2210, -0.0215] |
| 14 | [-0.0196, 0.2056] |
| 15 | [0.2212, 0.5017] |
| 16 | [0.5212, 0.7573] |
| 17 | [0.7745, 0.9446] |
| 18 | [0.9460, 1.0987] |
| 19 | [1.1005, 1.2774] |
| 20 | [1.3039, 1.5243] |
| 21 | [1.5413, 1.7098] |
| 22 | [1.7245, 1.8674] |
| 23 | [1.9008, 2.0266] |
| 24 | [2.0719, 2.2223] |
| 25 | [2.2682, 2.4397] |
| 26 | [2.4494, 2.7607] |

## 5.1 The construction of pattern moving space

In this simulation, the pattern moving space is constructed by fuzzy c-means (FCM) clustering algorithm, and the value of CPMS is 26. The dimension reduced operating condition data is clustered by FCM. The interval number constituted by the maximum and minimum values of data belonging to a certain class is used as the interval number representation of the pattern class. Table 1 presents the interval number representation of each pattern class in the pattern moving space, and each pattern class is represented (measured) by an interval number. The "points" in Fig. 2 represent the operating condition pattern. The "vertical lines" in the figure represent the pattern class variable time series corresponding to the operating condition pattern. $I_i^L$ and $I_i^U$ represent the maximum and minimum value of data belonging to class $i$, respectively.

## 5.2 The moving pattern-based modeling using the proposed interval ARX

In this simulation, the input of the model is ignition temperature $u(k)$, which is normalized by the zero-mean, and is illustrated in Fig. 3.

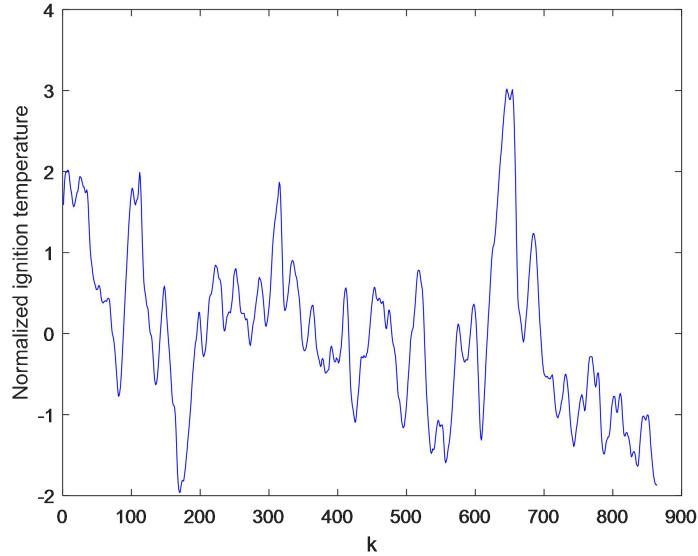

Fig. 3 The normalized ignition temperature.

The moving pattern-based model using the proposed IARX is as follows:

$$d\tilde{x}(k) = (p_{0c}, p_{0r}) + \sum_{j=1}^{3} dx(k-j) \circ \begin{bmatrix} p_j^1 \\ p_j^2 \end{bmatrix} + (p_{4c}, p_{4r})u(k-1) \quad (24)$$

Here, $d\tilde{x}(k)$ is the preliminary prediction output at time $k$, and $u(k)$ is the input. By formula (18), (19), the identified parameters are as follows:

$$\begin{aligned}
\mathbf{A}^T &= [p_{0c}, p_1^1, p_2^1, p_3^1, p_{4c}] \\
&= [0.0055,\ 1.2369,\ 0.0356,\ -0.2880,\ -0.0085] \\
\mathbf{C}^T &= [p_{0r}, p_1^2, p_2^2, p_3^2, p_{4r}] \\
&= [0.0124,\ 0.8898,\ 0.0000,\ 0.0000,\ 0.0018]
\end{aligned}$$

The preliminary prediction outputs obtained by the proposed method are illustrated in Fig. 4, Fig. 6, and Fig. 8, respectively. In these figures, the two solid blue lines represent the upper bound and lower bound of the preliminary prediction outputs, respectively, and the magenta vertical segments represent the pattern class variables time series. The final prediction outputs obtained by the proposed method are illustrated in Fig. 5, Fig. 7, and Fig. 9, respectively. In these figures, the two solid blue lines represent the upper bound and lower bound of the final prediction outputs, and the magenta vertical segment represents the time series of pattern class variables. By comparing Fig. 6 and Fig. 7, we can see in the period of K=639-651, the preliminary prediction output values are continuously changing, while the final prediction output values are discrete. This is because the preliminary prediction outputs need to be classified into a certain pattern class which is belong to the pattern moving space, and the interval number representation (metric) of this pattern class is used as the final prediction output.

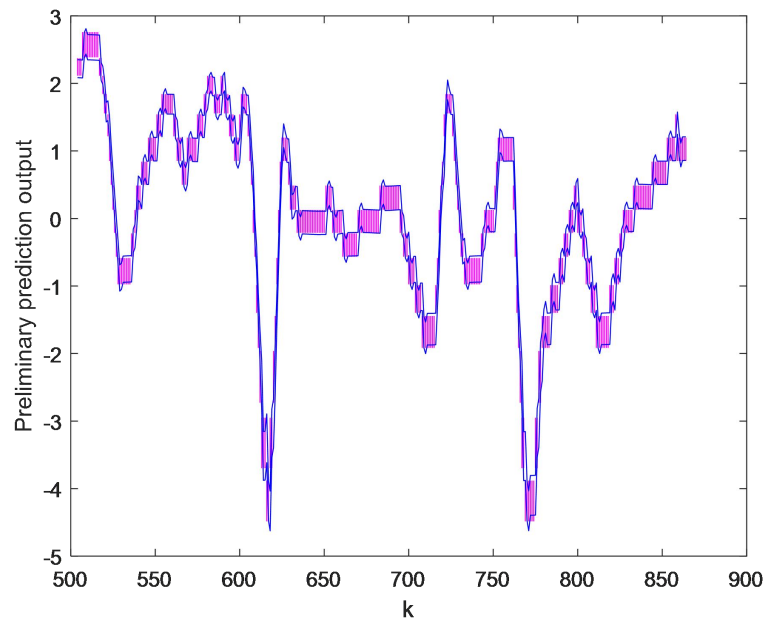

Fig. 4 The preliminary prediction output by the proposed model (CPMS=16).

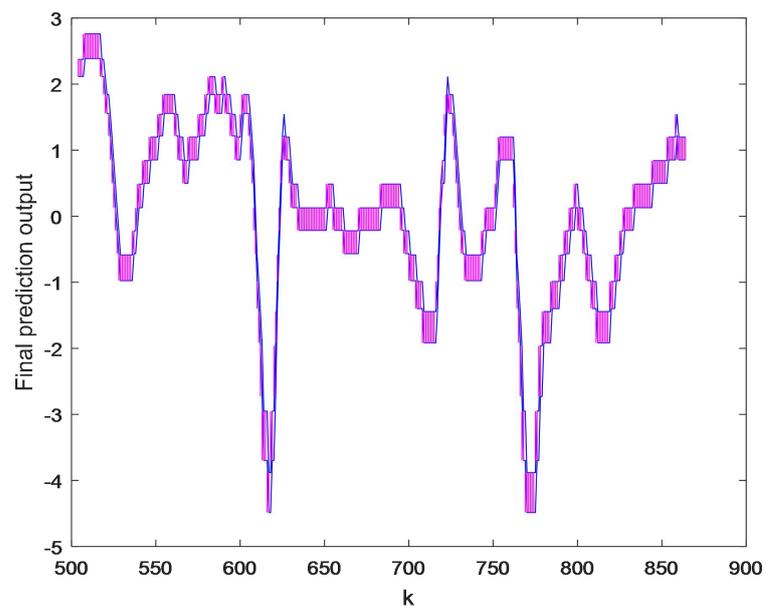

Fig. 5 The final prediction output by the proposed model (CPMS =16).

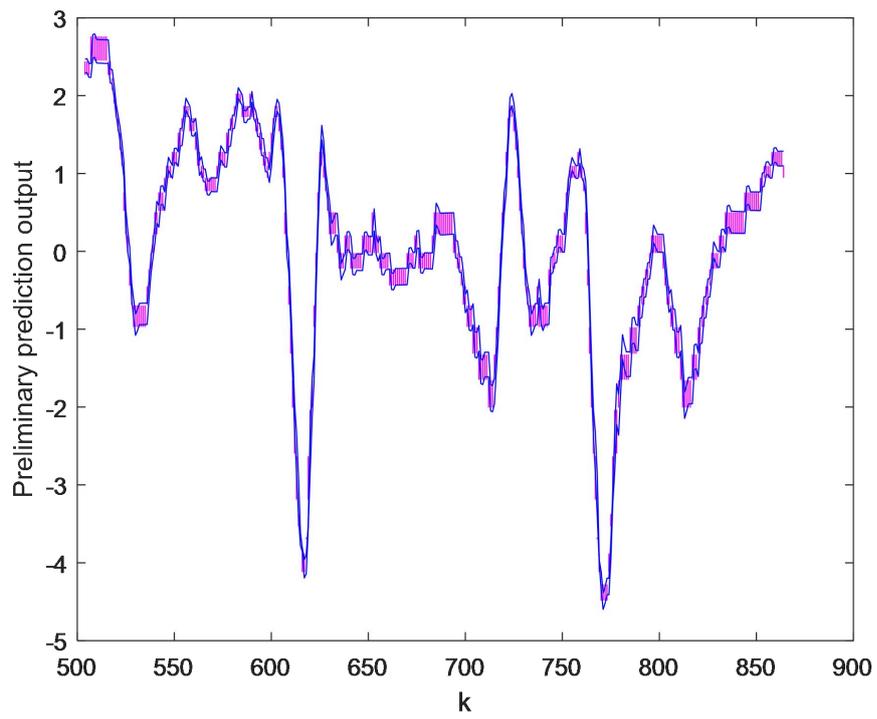

Fig. 6 The preliminary prediction output by the proposed model (CPMS =26).

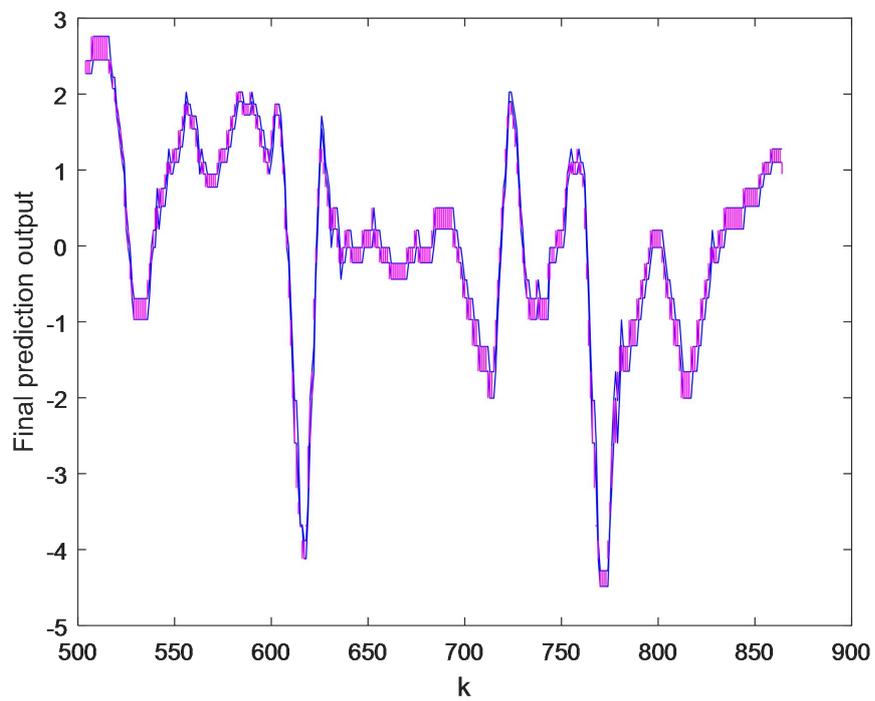

Fig. 7 The final prediction output by the proposed model (CPMS =26).

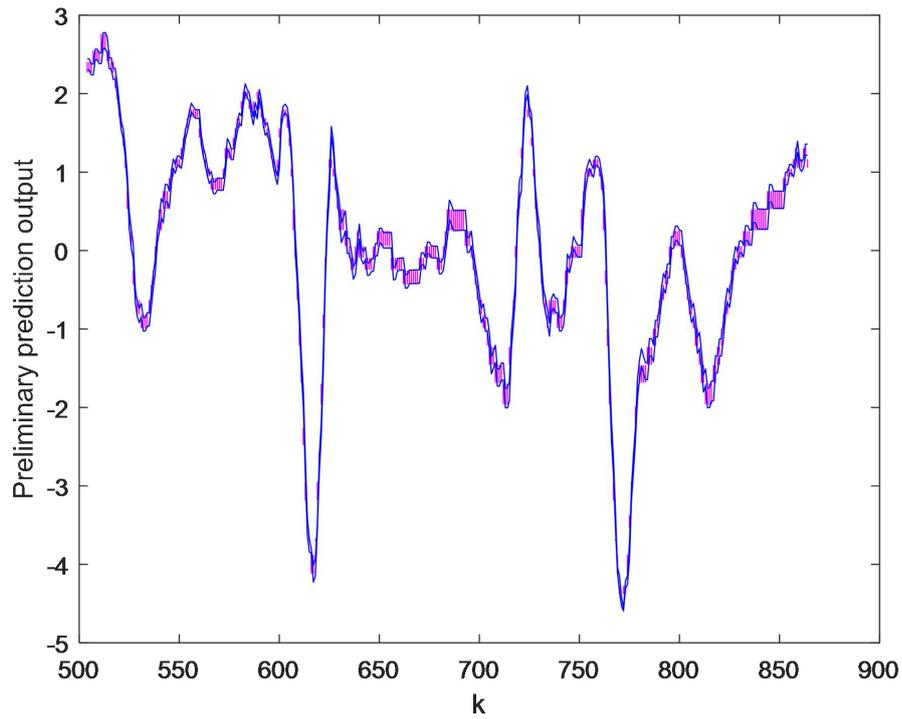

Fig. 8 The preliminary prediction output by the proposed model (CPMS =36).

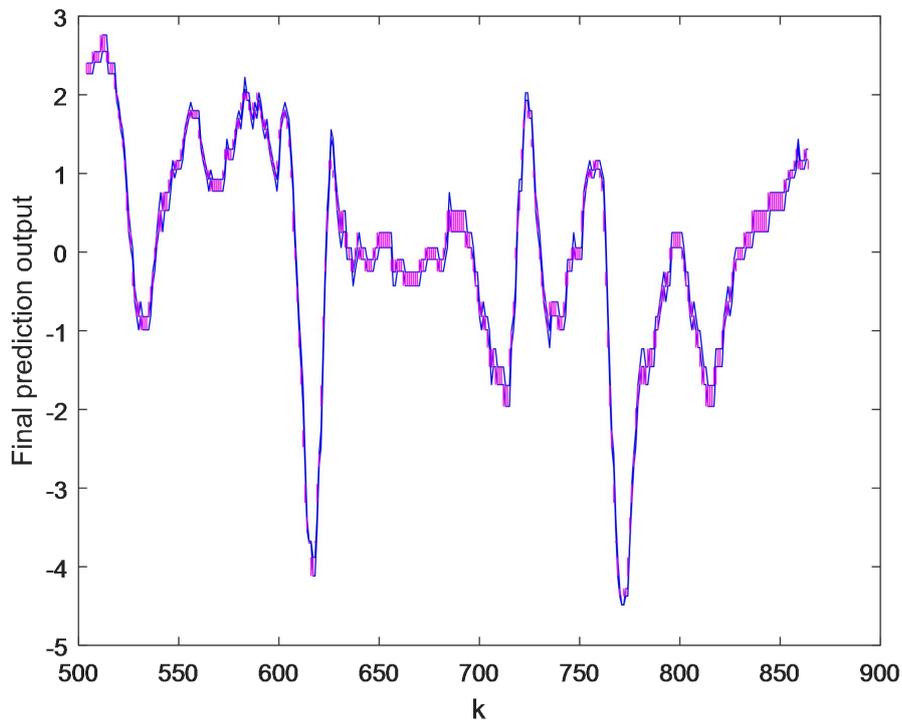

Fig. 9 The final prediction output by the proposed model (CPMS =36).

The comparisons of the performance between the proposed algorithm and that in [7] are shown in Table2, Table 3, Table 4 when the CPMS is 16, 26 and 36, respectively. It is found that the $\text{RMSE}_{d\tilde{x}^H}$, $\text{RMSE}_{d\tilde{x}^L}$, $\text{RMSE}_{d\hat{x}^H}$ and $\text{RMSE}_{d\hat{x}^L}$, which are obtained by the proposed model, are smaller than the corresponding results obtained by the model in [7] respectively when the CPMS is 26. When the CPMS is 16 or 36, the $\text{RMSE}_{d\tilde{x}^H}$ and $\text{RMSE}_{d\tilde{x}^L}$ are smaller than the corresponding results obtained in [7], and the $\text{RMSE}_{d\hat{x}^H}$ and $\text{RMSE}_{d\hat{x}^L}$ are equal to that obtained in [7].

Table II Comparison of the performance between the algorithm in this paper and that in [7] (CPMS=16)

|  | $\text{RMSE}_{d\tilde{x}^H}$ | $\text{RMSE}_{d\tilde{x}^L}$ | $\text{RMSE}_{d\hat{x}^H}$ | $\text{RMSE}_{d\hat{x}^L}$ |
|---|---|---|---|---|
| The algorithm in [7] | 0.2335 | 0.2398 | 0.2488 | 0.2572 |
| The proposed algorithm | **0.2302** | **0.2371** | **0.2488** | **0.2572** |

Table III Comparison of the performance between the algorithm in this paper and that in [7] (CPMS=26)

|  | $\text{RMSE}_{d\tilde{x}^H}$ | $\text{RMSE}_{d\tilde{x}^L}$ | $\text{RMSE}_{d\hat{x}^H}$ | $\text{RMSE}_{d\hat{x}^L}$ |
|---|---|---|---|---|
| The algorithm in [7] | 0.1763 | 0.1782 | 0.1912 | 0.1897 |
| The proposed algorithm | **0.1761** | **0.1729** | **0.1900** | **0.1873** |

Table IV Comparison of the performance between the algorithm in this paper and that in [7] (CPMS=36)

|  | $\text{RMSE}_{d\tilde{x}^H}$ | $\text{RMSE}_{d\tilde{x}^L}$ | $\text{RMSE}_{d\hat{x}^H}$ | $\text{RMSE}_{d\hat{x}^L}$ |
|---|---|---|---|---|
| The algorithm in [7] | 0.1520 | 0.1523 | 0.1607 | 0.1638 |
| The proposed algorithm | **0.1503** | **0.1516** | **0.1607** | **0.1638** |

**5.3 The relationship between the RMSE of the proposed model and the CPMS**

For further study the influence of the CPMS on the RMSE of the proposed model, the RMSE of preliminary prediction outputs corresponding to different values of CPMS is illustrated in Fig.10-11. In this simulation, the values of CPMS vary from 16 to 36. The RMSE of final prediction output corresponding to different values of CPMS is illustrated in Fig.12-13.

From Fig.10-11, with the values of CPMS varying from 16 to 36, the $\text{RMSE}_{d\tilde{x}^H}$ and $\text{RMSE}_{d\tilde{x}^L}$ of the proposed model are all smaller than the corresponding results in [7]. From Fig.12-13, it is found the results obtained in this paper are significantly smaller than those in [7] at individual point (e.g., when the value of CPMS is 26). While, at other points, the $\text{RMSE}_{d\hat{x}^H}$ and $\text{RMSE}_{d\hat{x}^L}$ of the proposed model almost completely coincide with those given in [7]. In addition, it is found that from Fig. 10-13 the root mean square error shows an overall downward trend with the increase of the number of pattern class.

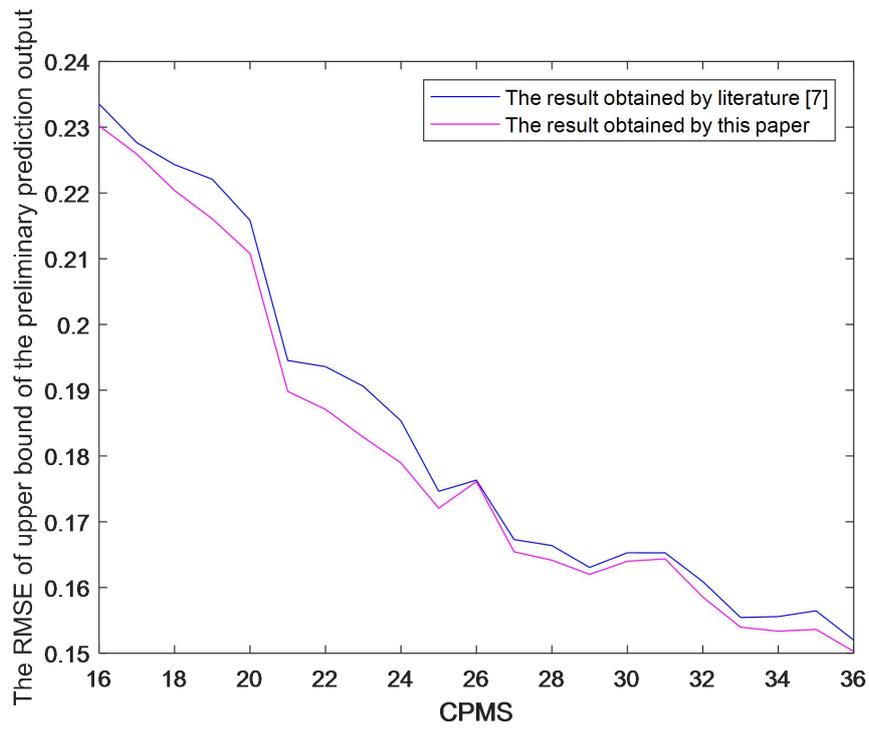

Fig.10 The $\text{RMSE}_{d\tilde{x}^H}$ corresponding to different values of CPMS.

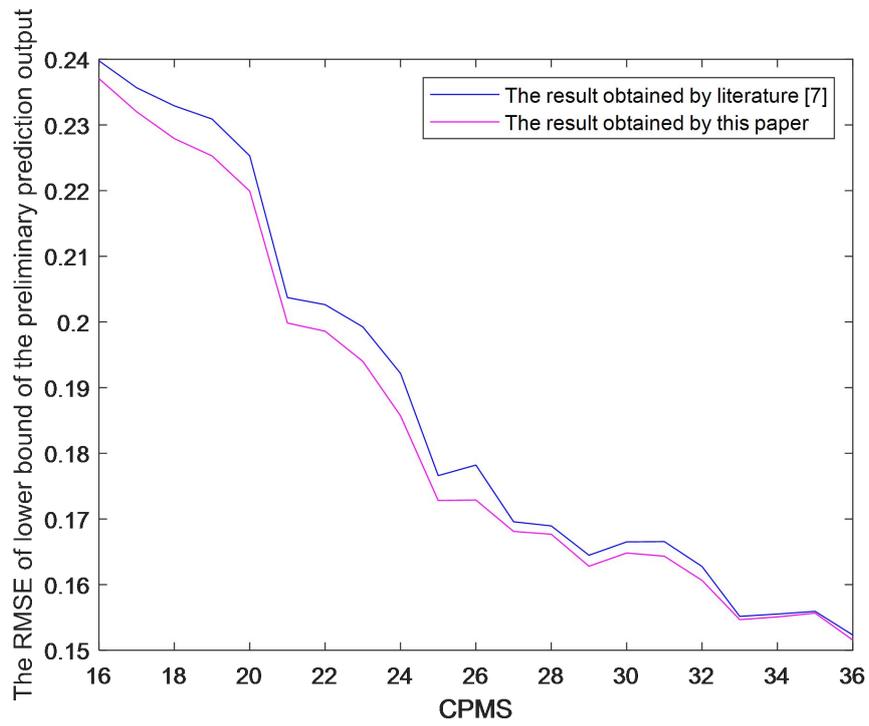

Fig.11 The $\text{RMSE}_{d\tilde{x}^L}$ corresponding to different values of CPMS.

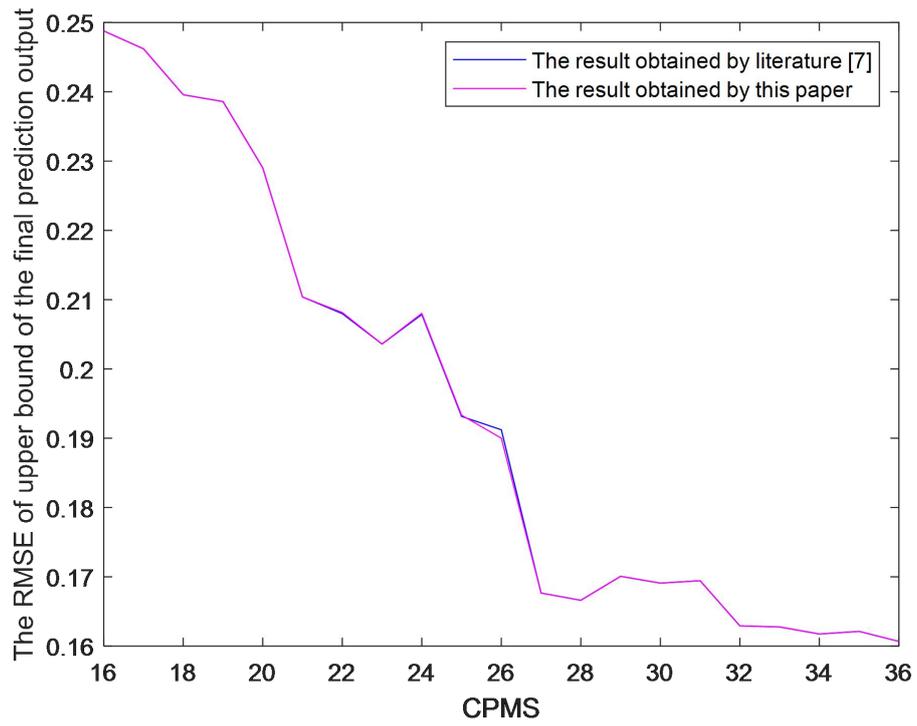

Fig.12 The $\text{RMSE}_{d\hat{x}^H}$ corresponding to different values of CPMS.

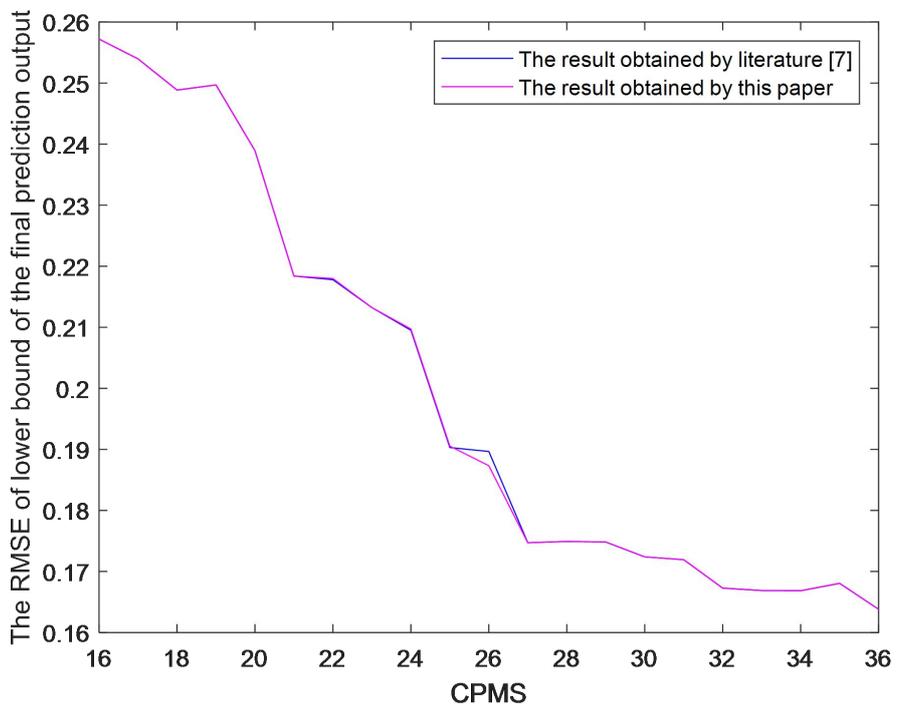

Fig.13 The $\text{RMSE}_{d\hat{x}^L}$ corresponding to different values of CPMS

## 5.4 The robustness of moving pattern-based modeling using the new type of interval ARX model

In this section, the robustness of moving pattern-based modeling using the new type of interval ARX model is discussed. Moving pattern-based modeling is essentially the construction of a classifier. The parameter changes of the preliminary prediction model lead to certain changes in its output. However, as can be seen from Formula 1, the output of the preliminary prediction model goes through a classification process to obtain the final forecast output. The changes in the output of the preliminary prediction model are "digested" in the classification process, and the correct classification can still be obtained eventually, that is: although the parameters of the preliminary prediction model have changed to some extent, the RMSE of final prediction output remains the same. Of course, the ability to "digest" the output changes of the preliminary prediction model through classification is not infinite. When the output changes of the preliminary prediction model exceed a certain range, it will lead to classification errors, which will further affect the RMSE of final prediction output.

In this simulation, by adding uniformly distributed random numbers to initial parameters $C$ of the model, the varied parameters are simulated. In table V, the model initial parameters $C$ and varied parameters $C_1$ are as follows:

$$C^T = [p_{0r}, p_1^2, p_2^2, p_3^2, p_{4r}]$$
$$= [0.0124, \ 0.8898, \ 0.0000, \ 0.0000, \ 0.0018]$$

$$C_1^T = [p_{0r}, p_1^2, p_2^2, p_3^2, p_{4r}]$$
$$= [0.0322, \ 0.9097, \ 0.0272, \ 0.0604, \ 0.0034]$$

Table V The robustness of the proposed modeling method

| Model parameters | $\text{RMSE}_{d\tilde{x}^H}$ | $\text{RMSE}_{d\tilde{x}^L}$ | $\text{RMSE}_{d\hat{x}^H}$ | $\text{RMSE}_{d\hat{x}^L}$ |
|---|---|---|---|---|
| $C$ | 0.1761 | 0.1729 | 0.1900 | 0.1873 |
| $C_1$ | 0.1813 | 0.1749 | 0.1900 | 0.1873 |

From table V, it is found that although the upper bound and lower bound RMSE of the preliminary prediction output increased after the parameters changed from $C$ to $C_1$, the upper and lower bound RMSE of the final prediction output did not change. This proves that changes in the preliminary prediction output caused by changes in preliminary prediction model parameters can be "digested" by the classification process. It indicates the robustness of the proposed modeling method in this paper.

## 6 Conclusions

In this paper, a new type of interval ARX model is proposed, and is applied to moving pattern-based modeling. A new interval operation is defined, namely, the "multiplication" operation between an interval number and a real matrix, and the multiplied result is an interval matrix. In particular, $m=1$ in the definition, the defined operation is simplified to the operation we defined earlier in [18]. Different from traditional ARX models, the output of the proposed new type of interval ARX model based on the defined operation is an interval number. For better characterizing pattern moving space, the "cardinality" of pattern moving space is defined. CPMS characterize the granularity of pattern moving space. The larger the value of CPMS is, the more finely the original dynamic system is characterized. To verify the validity of the proposed model, it is applied to the sintering process. The simulation results show that the upper and lower bound root mean square error of the preliminary prediction output of the proposed model are smaller than the corresponding results in [7], respectively. Moreover, the upper bound root mean square error curves, lower bound root mean square error curves of the final prediction output of the proposed model are significantly smaller than those in [7] at individual point (e.g., when the value of CPMS is 26). While, at other points, they almost completely coincide with those given in [7]. In addition, the changes in the output of the preliminary prediction model could be "digested" in the classification process, the proposed modeling method is robust to variation in model parameters.

In future work, the application of the proposed model for interval time series forecasting, robust control and fault detection will be researched.

**Acknowledgments**
The author wishes to thank Professor Guo Xianggui for valuable discussions.